\documentclass[12pt,preprint]{aastex}
\usepackage{epsf}
\newcommand{\etal}{{\it et al.}}
\newcommand{\beq}{\begin{equation}}
\newcommand{\eeq}{\end{equation}}
\newcommand{\ben}{\begin{eqnarray}}
\newcommand{\een}{\end{eqnarray}}
\newcommand{\br}{{\bf r}}

\newcommand{\bu}{{\bf u}}

\newcommand{\rhog}{\rho_{g}}

\newcommand{\Pg}{P_{g}}
\newcommand{\ep}{\epsilon}
\begin{document}
\title{Modeling Intra-Cluster Gas in Triaxial Dark Halos : \\ 
An Analytic Approach}
\author{\sc Jounghun Lee and Yasushi Suto}
\affil{Department of Physics, The University of Tokyo, 
 Tokyo 113-0033, Japan}
\email{lee@utap.phys.s.u-tokyo.ac.jp, suto@phys.s.u-tokyo.ac.jp}
\received{2002 August 15}
\accepted{2002 ???}
\begin{abstract} 
We present the first physical model for the non-spherical intra-cluster
gas distribution in hydrostatic equilibrium under the gravity of
triaxial dark matter halos.  Adopting the concentric triaxial
density profiles of the dark halos with constant axis ratios proposed by
Jing \& Suto (2002), we derive an analytical expression for the 
triaxial halo potential on the basis of the perturbation theory, 
and find the hydrostatic solutions for the gas density and temperature 
profiles both in isothermal and polytropic equations of state. 
The resulting iso-potential surfaces are well approximated by triaxial 
ellipsoids with the eccentricities dependent on the radial distance.   
We also find a formula for the eccentricity ratio between the 
intra-cluster gas and the underlying dark halo. Our results allow one  
to determine the shapes of the underlying dark halos from the observed   
intra-cluster gas through the X-ray and/or the Sunyaev-Zel'dovich effects 
clusters. 
\end{abstract} 
\section{INTRODUCTION}

While clusters of galaxies are regarded as one of the most important
cosmological probes, the conventional modeling of the intra-cluster gas
is very approximate at best. The most popular and traditional
description is the spherical isothermal $\beta$ model, which 
has been claimed as a good empirical description for the observed
intra-cluster gas, and in fact has been widely
used in practically almost all statistical analyses of galaxy clusters
including mass, temperature and luminosity functions, and the Hubble
constant (or more precisely the angular diameter distance) measurement
via the Sunyaev-Zel'dovich (SZ) effect 
\citep[e.g.,][for a review]{birkinshaw99}.

The situation is changing rapidly both observationally and theoretically; 
recent X-ray satellites such as {\it Chandra} and {\it XMM-Newton} 
have greatly improved the angular-resolution and the statistical reliability 
of X-ray maps of galaxy clusters, and often detected the significant 
departure from the isothermality and the non-sphericity of the intra-cluster 
gas.  \citet{buote02}, for instance, reported evidence from the {\it
Chandra} observation for a flattened triaxial dark matter halo around 
the elliptical galaxy NGC720 \citep[see also, ][]{RK98}.  
In the radio band the bolometer array technique
improved the angular resolution of the cluster SZ map. 
One successful example is the discovery of substructure of the most
luminous X-ray cluster RX J1347-1145 by \cite{komatsu01}, 
which was subsequently confirmed with {\it Chandra} (Allen et
al. 2002). 

The theoretical understanding of dark matter halos is also advanced 
significantly. The spherical averaged density profiles of dark halos 
are known to be well approximated by a sequence of universal density profiles
\citep{nav-etal96,nav-etal97,fuk-mak97,moo-etal98,jingsuto00}.  
The importance of non-spherical effects on the mass function of dark halos
was first raised by \citet{mon95}, and then studied analytically 
\citep{aud-etal97,LS98}, and in detail with numerical simulations 
\citep{LS99,sheth99,jenkins01}. 
More importantly, the systematic effect of the non-sphericity of
intra-cluster gas  on the estimate of the Hubble constant from the SZ 
observations is well recognized 
\citep{BHA91,inagaki95,yoshikawa98,HB98,fox-pen02}.
In particular the recent work of Jing \& Suto (2002) showed that the
simulated halos are well approximated by a sequence of the concentric
triaxial distribution with their axis directions being fairly aligned.
This naturally opens a possibility to compute the hydrostatic
equilibrium solution of the intra-cluster gas under the gravity of the
triaxial dark matter halos. Generalizing the prescription of
Makino, Sasaki \& Suto (1998) and Suto, Sasaki, \& Makino (1998) who
obtained the hydrostatic equilibrium solution for the spherical halo
profiles, we find a series of analytical solutions for gas embedded in
the triaxial halo profiles. In the present paper, we focus on the
mathematical aspects of the problem, and their astrophysical
applications (weak/strong lensing, SZ and X-ray observations) will be
described elsewhere.

The rest of the paper is organized as follows: In $\S 2$ we briefly
outline the procedure to compute gas and temperature profiles in
hydrostatic equilibrium with gravitational potential of dark matter
halos. In $\S 3$ we present the analytical derivation of the triaxial halo
potential using the perturbation theory, and describe the shapes of 
the iso-potential surfaces. The analytical and numerical results for the 
gas density and temperature profiles are summarized in $\S 4$, and $\S 5$ is
devoted to conclusions and discussion. In Appendix, we provide the
detailed analytical expressions of our perturbative results. 

\section{DISTRIBUTION OF INTRA-CLUSTER GAS IN HYDROSTATIC EQUILIBRIUM}

If the intra-cluster gas embedded in the dark matter halo is 
in hydrostatic equilibrium, its distribution is described by 
\beq 
\frac{1}{\rhog}\nabla\Pg = -\nabla\Phi, 
\label{eqn:hydrostatic}
\eeq 
where $\Phi$ is the gravitational potential of the system, and $\rhog$
and $\Pg$ represent the density and the pressure of the intra-cluster
gas, respectively.  
In what follows, we neglect the contribution of the gas and
stellar masses, and assume that the gravitational potential of the 
total system is well approximated by that of the halo \citep{sut-etal98}.

Once the equation of state for the intra-cluster gas and the density
profile of the dark matter halo are given, then the density and 
temperature profiles of intra-cluster gas are obtained by solving equation 
(\ref{eqn:hydrostatic}).  Although the intra-cluster gas is
commonly assumed to be isothermal, it has been recently claimed that the
intra-cluster gas might be better described as polytropic 
\citep{mar-etal98,kom-sel01}.  
Thus we consider both isothermal and polytropic cases.

Consider first the isothermal gas with an equation of state: 
\begin{eqnarray}
 P_{g} = K\rho_{g}, \qquad K \equiv \frac{k_B T_{\rm g}}{\mu m_p} ,
\end{eqnarray}
where $T_{\rm g}$, $k_{B}$, $\mu$, and $m_{p}$ denote the (constant) gas
temperature, the Boltzmann constant, the mean molecular weight, and the
proton mass, respectively.  In this case the hydrostatic equilibrium
equation (\ref{eqn:hydrostatic}) is easily solved to yield
\beq
\frac{\rhog}{\rho_{g0}} = \exp\left[-\frac{1}{K}(\Phi - \Phi_{0})\right],  
\label{eqn:iso}
\eeq
where $\Phi_{0}$ is an integration constant that can be fixed by the 
condition of $\rho_{g}(r=0) = \rho_{g0}$. Note that the value of 
$\Phi_{0}$ depends on the value of $K$.  

Turn next to the polytropic model with an equation of state: 
\begin{eqnarray}
\label{eqn:pgas}
P_{g} = K_0 \rhog^{\gamma}, \qquad 
K_0 \equiv \frac{k_B T_{\rm g0}}{\mu m_p} ,
\end{eqnarray}
where $\gamma$($\ne 1$) is the polytropic index, and $T_{g0}$ is the 
value of the gas temperature at $r = 0$. In this polytropic model, 
the gas temperature is not constant but proportional to
$\rho_{g}^{\gamma -1}$. So, in this case, equation
(\ref{eqn:hydrostatic}) reduces to 
\begin{eqnarray}
\nabla (\rho^{\gamma -1}) = -\frac{\gamma - 1}{K_0 \gamma}\nabla\Phi, 
\end{eqnarray}
whose solution is derived as
\beq
\label{eqn:pol} 
\frac{T_{g}}{T_{g0}} = \frac{1-\gamma }{K_0\gamma}(\Phi - \Phi_{0}),
\qquad
\frac{\rhog}{\rho_{g0}} = 
\left[\frac{1-\gamma }{K_0\gamma}(\Phi - \Phi_{0})\right]^{1/(\gamma-1)}.  
\eeq
Thus, the general density and temperature profiles of
intra-cluster gas in both isothermal and polytropic cases can be 
straightforwardly obtained in terms of the gravitational potential of
the underlying halo.  
Furthermore, equations (\ref{eqn:iso}) and (\ref{eqn:pol}) also imply 
that the iso-potential surfaces of the triaxial dark halo coincide with the
iso-density surfaces of the intra-cluster gas.  This is simply a direct
consequence of {\it X-ray shape theorem} \citep{buote94,buote02}; the
hydrostatic equilibrium equation (\ref{eqn:hydrostatic}) yields
\beq
\nabla\Pg \times \nabla\Phi = \nabla\rhog \times \nabla\Phi = 0 .
\eeq
Moreover, with the polytropic equation of state (eq. [\ref{eqn:pgas}]), 
the gas temperature also satisfies
\beq
\nabla T_{g}  \times \nabla\Phi =  0 .
\eeq
Therefore, the gas density, temperature, and pressure are constant on 
the iso-potential surfaces at the same time. In consequence,  
the gravitational potential is the central quantity for the hydrostatic 
gas model. In the following section we derive an analytical expression 
for the triaxial dark halo  potential, and study the general properties  
of the resulting iso-potential surfaces.

\section{GRAVITATIONAL POTENTIAL OF TRIAXIAL DARK HALOS}

\subsection{Triaxial Density Profile of Dark Halos}
In what follows, we adopt the following concentric triaxial density
profile for the dark matter halos \citep{jingsuto02}:
\beq
\label{eqn:den}
\rho(R) = \frac{\delta_{c}\rho_{\rm crit}}{\left(R/R_0\right)^{\alpha}
\left(1 + R/R_0\right)^{3-\alpha}} ,
\eeq
where $R_{0}$ is the scale length, $\delta_{c}$ is the dimensionless 
characteristic density contrast with respect to the critical density 
$\rho_{\rm crit}$ of the universe at the present epoch, and 
$\alpha$ represents the inner slope of the density profile. 
Note that equation (\ref{eqn:den}) is identical to the one which
describes the spherical density profile 
\citep{nav-etal97,fuk-mak97,moo-etal98} except
that the spherical radius $r$ is replaced by the the major axis length
$R$ of the iso-density surfaces :
\begin{eqnarray}
\label{eq:isodensity}
R^2= a^{2}\left(\frac{x^2}{a^2} + 
\frac{y^2}{b^2} + \frac{z^2}{c^2}\right), \qquad (a \ge b \ge c).
\end{eqnarray}
The N-body simulations by Jing \& Suto (2002) imply that the best-fit
values of $\alpha$ are $\alpha = 1$ and $\alpha = 3/2$ for cluster and 
galactic-scale halos, respectively. Thus we pay particular attention 
to these two cases.   

We quantify the ellipsoidal shape of a {\it halo} iso-density surface by 
defining the two eccentricities: 
\beq
e_b \equiv \sqrt{1-\left(\frac{b}{a}\right)^2} ,
\quad
e_c \equiv \sqrt{1-\left(\frac{c}{a}\right)^2} , 
\label{eqn:e}
\eeq 
and $a \ge b \ge c$ implies $e_{b} \le e_{c}$ . The values of
 $e^{2}_{\sigma}$ ($\sigma = b, c$) measure the degree of the deviation 
of the ellipsoidal iso-density surfaces from the spherical ones along 
the corresponding principal axis direction.  

\subsection{Perturbative Expansion of the Triaxial Halo Potential}

The gravitational potential of a dark halo with the triaxial 
density profile (eq.[\ref{eqn:den}]) can be written as \citep{bin-tre87}
\begin{eqnarray}
\Phi(\br) &=& -\pi G \left( \frac{bc}{a}\right)
\int^{\infty}_{0}\frac{[\psi(\infty)-\psi(m)]}
{\sqrt{(\tau + a^2)(\tau + b^2)(\tau + c^2)}}d\tau, \label{eqn:phi1}\\
\psi(m) &=& 2\int_{0}^{m}\rho(R)RdR,\label{eqn:psi}\\ 
m^{2} &=& a^{2}\left(\frac{x^{2}}{a^2 + \tau} + 
\frac{y^{2}}{b^{2} + \tau} + \frac{z^{2}}{c^{2} + \tau}\right), 
\label{eqn:m}
\end{eqnarray}
where $\br=(x,y,z)=r(\sin\theta\cos\phi,\sin\theta\sin\phi,\cos\theta)$, 
and $\tau$ labels the iso-potential surfaces on which $m = m(\br,\tau)$ 
is constant. While numerical integration is required in general to obtain the
triaxial gravitational potential with equations (\ref{eqn:den}) to 
(\ref{eqn:m}), small eccentricities ($e_{b}^{2} \le e_{c}^{2} \ll 1$) 
cases enable us to approach this problem analytically with the perturbative 
expansion. 

The theory for the ellipsoidal perturbation of the spherical systems has 
a long history \citep[e.g., see][and references therein]{jef76}. 
In performing the ellipsoidal perturbation, there are two different 
ways to arrange the perturbed terms. 
One is the equal-volume approach: the perturbed 
ellipsoids have the same volumes as the unperturbed original spheres. 
The other is the equal-length approach: the perturbed 
ellipsoids and the unperturbed spheres have the same length scales. 
For instance, one can force the major axis lengths of the perturbed 
ellipsoids to be same as the radii of the original spheres. 
The two approaches differ only in arranging the perturbed ellipsoidal
terms, and the perturbation theory is basically the same either way.  
Though the equal-volume approach is more conventional, we take the 
equal-length approach here, for which the reason is as follows.

We develop the perturbation theory here mainly to provide an ellipsoidal 
correction to the conventional spherical modeling of the observed
clusters. What is observationally measurable, however, is not the
volumes but the length scales of the clusters. For example, in  
SZ observations, the cluster length scales in the direction
perpendicular to the line-of-sight are measured. In the standard
spherical model, this tangential length scales are taken as the
spherical radii of the clusters. In ellipsoidal model, however, 
the same tangential length scales are no longer same as the
line-of-sight length. Thus, in the two models, the volumes of the
clusters are not the same. 
Thus, given the future application of the perturbation theory 
to the observed clusters, we perform the perturbative expansion of the
gravitational potential in such a way that the perturbed ellipsoids 
have the same length scales (here, the major axis lengths) as the 
radii of the spheres but do not have the same volumes. Derivation of the 
perturbative expansion of $\Phi(\br)$ up to the first order of 
$e_{\sigma}^{2}$ in this way proceeds as follows. 

First, $\psi(m)$ defined in equation (\ref{eqn:psi}) for the 
given triaxial density profile $\rho(R)$ in equation (\ref{eqn:den}) 
is analytically given as 
\beq
\psi(m) = \frac{2R_{0}^{2}\delta_{c}\rho_{\rm crit}}{2-\alpha}
\left(1 + \frac{R_{0}}{m}\right)^{\alpha - 2}.
\label{eqn:psi2}
\eeq
Expressing $b$ and $c$ in terms of $a$ as 
$b = a\sqrt{1-e_{b}^{2}}$ and $c = a\sqrt{1-e_{c}^{2}}$ by 
equation (\ref{eqn:e}), we expand $m$ in equation ($\ref{eqn:m}$) 
with respect to $e_{b}^{2}$ and $e_{c}^{2}$ to their its firstorder:  
\beq
m \approx m_{sp}\left(1 + \frac{m_{sp}^{2}}{r^{2}}
\frac{e_{b}^{2}\sin^{2}\theta\sin^{2}\phi + 
e_{c}^{2}\cos^{2}\theta}{2}\right), 
\label{eqn:m2}
\eeq
where $m_{sp} \equiv ar/\sqrt{a^{2}+\tau}$ is the value of $m$ 
for the spherical case ($e_{b} = e_{c} = 0$). 

Equations (\ref{eqn:psi2}) and $(\ref{eqn:m2})$ yield the first order 
approximation of $\psi(m)$ : 
\beq
\psi(m) \approx 2R_{0}^{2}\delta_{c}\rho_{\rm crit}
\left(1 + \frac{R_{0}}{m_{sp}}\right)^{\alpha -2}
\left(\frac{1}{2-\alpha} + \frac{R_{0}}{m_{sp}+R_{0}}\frac{m_{sp}^{2}}{r^{2}}
\frac{e_{b}^{2}\sin^{2}\theta\sin^{2}\phi + 
e_{c}^{2}\cos^{2}\theta}{2}\right). 
\label{eqn:psi3}
\eeq
Similarly, we approximate $\sqrt{(\tau + a^2)(\tau + b^2)(\tau + c^2)}$ 
and  $bc/a$ in equation (\ref{eqn:phi1}) as  
\beq
\sqrt{(\tau + a^2)(\tau + b^2)(\tau + c^2)} \approx 
\frac{m_{sp}^{3}}{a^{3}r^{3}}\left(1 + \frac{m_{sp}^{2}}{r^{2}}
\frac{e_{b}^{2}+e_{c}^{2}}{2}\right),\quad 
\frac{bc}{a} \approx a\left(1 - \frac{e_{b}^{2}+e_{c}^{2}}{2}\right). 
\label{eqn:min}
\eeq

Now that all the quantities in the integral of equation (\ref{eqn:phi1}) 
are expressed in terms of $m_{sp}$ as equations (\ref{eqn:m2}) to 
(\ref{eqn:min}), we change the integration variable from $\tau$ to 
$m_{sp}$, and  perform the integration of equation (\ref{eqn:phi1}) to 
obtain the following 1st-order approximation for $\Phi$:  
\ben
\label{eqn:gen}
\Phi(\bu) \approx C\left[ F_{1}(u) +
\frac{e_{b}^{2}+e_{c}^2}{2}F_{2}(u) 
 + \frac{e_{b}^{2}\sin^{2}\theta\sin^{2}\phi 
+  e_{c}^{2} \cos^{2}\theta}{2} F_{3}(u) \right],
\een
where $\bu \equiv \br/R_{0}$, and 
$C = 4\pi G\delta_{c}\rho_{\rm crit}R_{0}^{2}$, 
and the three functions, $F_{1}(u),F_{2}(u)$,and $F_{3}(u)$ are defined as
\ben
F_{1}(u) &\equiv& \frac{1}{\alpha-2}
\left[ 1 - \frac{1}{u}
\int_{0}^{u}\left(\frac{t}{t+1}\right)^{2-\alpha}dt
\right] , \label{eqn:f1}\\
F_{2}(u) &\equiv& \frac{1}{\alpha-2}
\left[ - \frac{2}{3}
+ \frac{1}{u}
\int_{0}^{u}\left(\frac{t}{t+1}\right)^{2-\alpha}dt
- \frac{1}{u^3}
\int_{0}^{u}\frac{t^{4-\alpha}}{(t+1)^{2-\alpha}}dt
\right] , \label{eqn:f2}\\
F_{3}(u) &\equiv& \frac{1}{u^3}
\int_{0}^{u}\frac{t^{4-\alpha}}{(t+1)^{3-\alpha}}dt . \label{eqn:f3}   
\een
Equation (\ref{eqn:gen}) is valid for any arbitrary value of $\alpha$. 
For the interesting values of $\alpha = 1$ and $\alpha = 3/2$, 
$F_{1}$, $F_{2}$ and $F_{3}$ can be written in terms of elementary 
functions (see Appendix A).
  
\begin{figure}
\begin{center}
\plotone{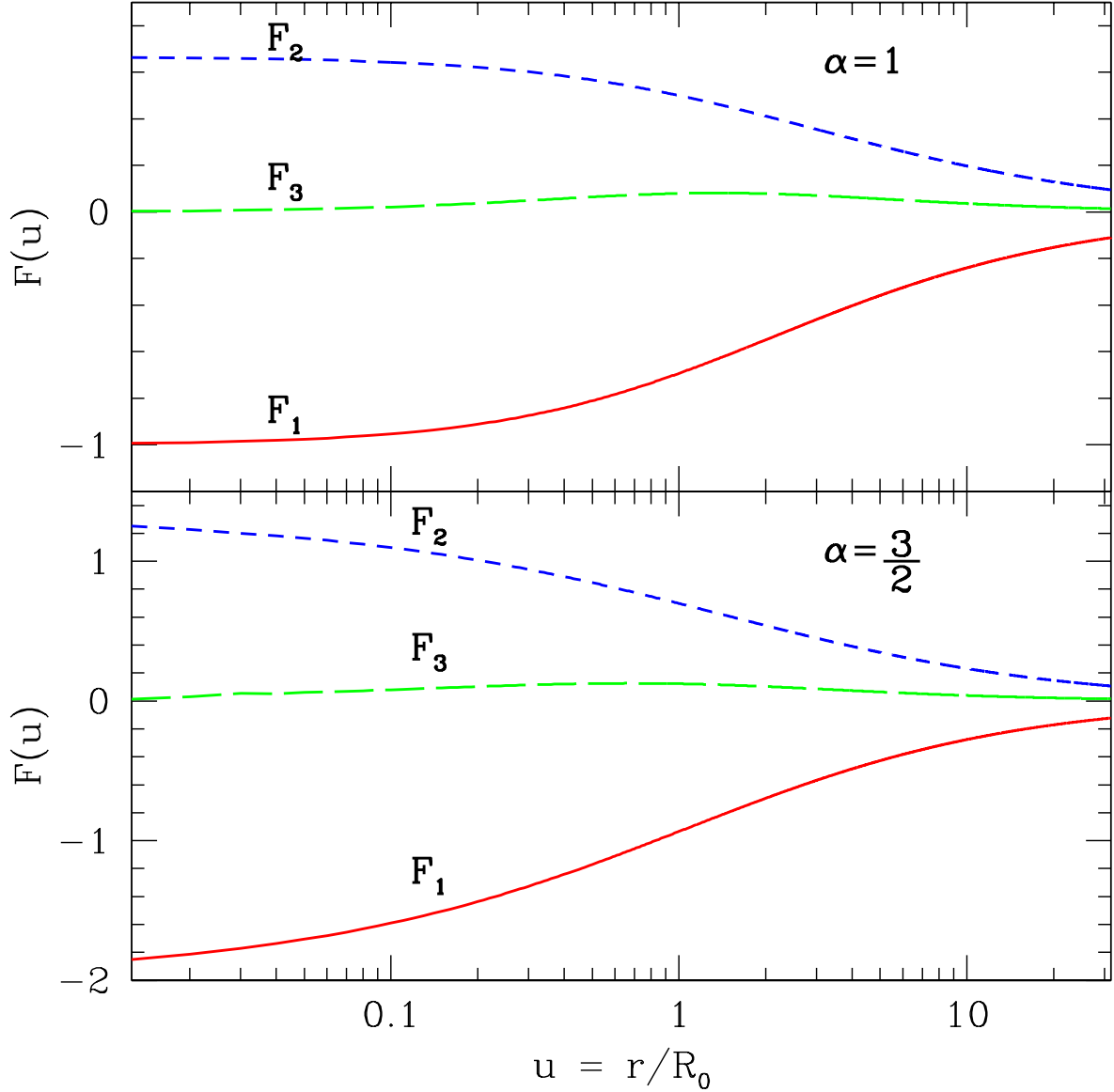}
\caption{The behavior of the three functions, $F_{1},F_{2}$ and $F_{3}$ 
(solid, dashed, and long-dashed lines, respectively) given in the 
perturbative result of the gravitational potential 
(eqs.[\ref{eqn:f1}],[\ref{eqn:f2}] and [\ref{eqn:f3}]). 
{\it Upper Panel}: it corresponds to the case where the inner slope 
of the halo density profile defined in equation (\ref{eqn:den}) 
has the value of $\alpha = 1$. 
{\it Lower Panel}: it corresponds to $\alpha = 3/2$. 
\label{fig:fs}}
\end{center}
\end{figure}

Here, $F_{1}(u)$ represents the spherical contribution to $\Phi(\bu)$, 
i.e., $CF_{1}(u) = 4\pi G\delta_{c}\rho_{\rm crit}R_{0}^{2}F_{1}(u)$ 
is nothing but the gravitational potential for the case of the spherical 
density profile \citep{mak-etal98,sut-etal98}. $F_{2}$ represents
another spherical contribution that has arisen due to the volume changes 
of the perturbed ellipsoidal density profiles from the spherical ones, 
while $F_{3}$ represents the non-spherical deviation of $\Phi(\bu)$ from 
the spherical potential.  If we take the equal-volume approach in 
performing the perturbation, then we will end up with exactlly same 
$F_{1}(u)$ and $F_{3}(u)$ but no $F_{2}$. 
Figure \ref{fig:fs} plots these functions in a wide range of the 
rescaled radius $u$ for $\alpha = 1$ and $\alpha = 3/2$. 
As shown, the magnitude of $F_{3}$ is an order of magnitude smaller 
than those of $F_{1}$ and $F_{2}$.  
Since the dependence of $\Phi(\bu)$ on $\theta$ and $\phi$ comes from 
the $F_{3}$-term (eq.[\ref{eqn:gen}]), it implies that $\Phi(\bu)$ is 
fairly insensitive to $\theta$ and $\phi$. The asymptotic limit of 
$F_{3}/F_{1}$ at $u \rightarrow \infty$ has been calculated to converge 
to zero,   which implies that the triaxial potential is almost 
indistinguishable from the spherical one at large distance ($u \gg 1$). 

\begin{figure}
\begin{center}
\plotone{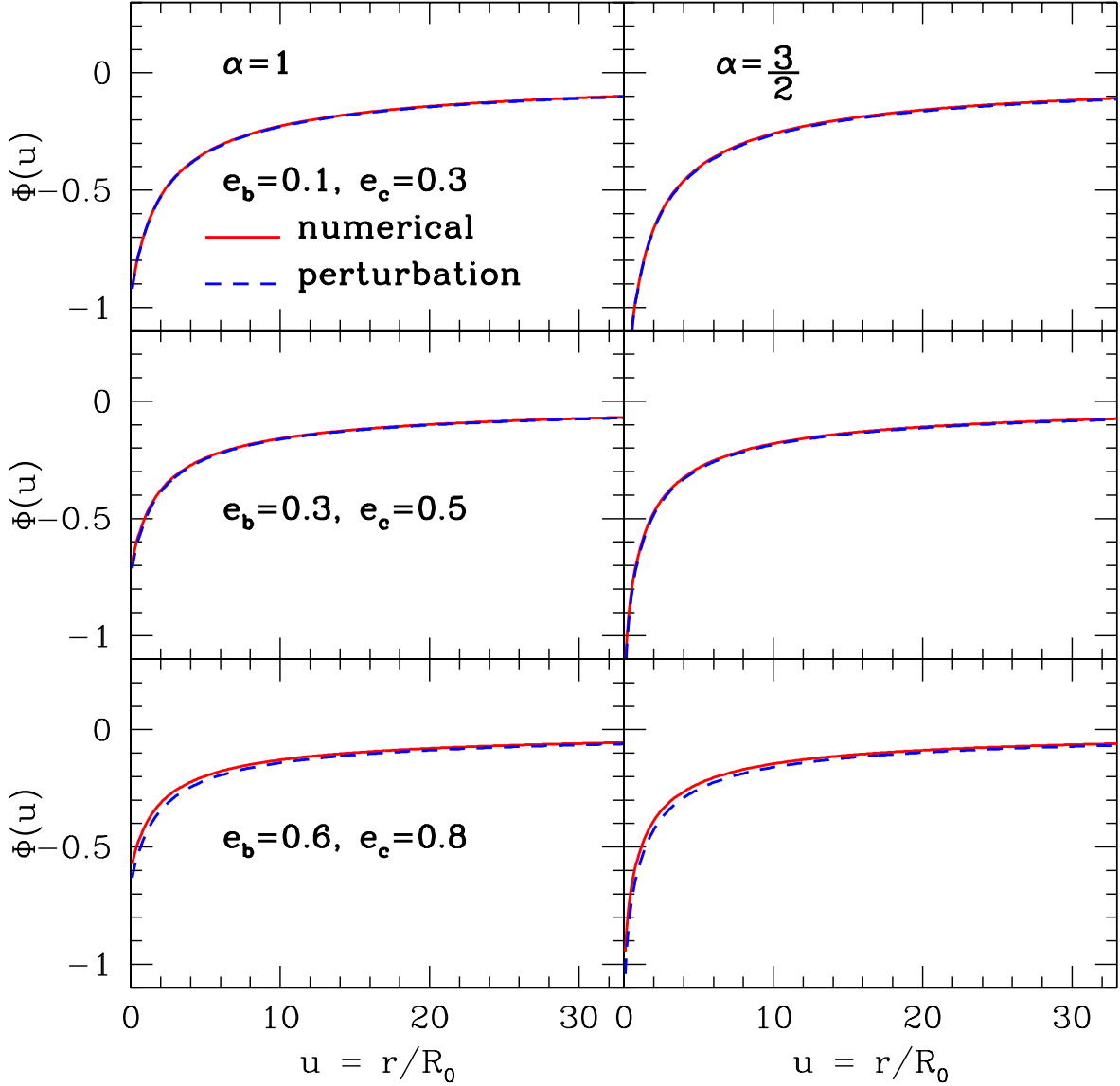}
\caption{Comparison of the numerical gravitational potentials (solid 
lines) with the perturbative results (dashed lines) 
on the equatorial plane of $\theta=\frac{\pi}{2}$ and $\phi=0$. 
{\it Top}: it correspond to the case where the underlying halo 
has the eccentricities, $e_{b} = 0.1, e_{c} = 0.3$.
{\it Middle}: $e_{b} = 0.3, e_{c} = 0.5$.
{\it Bottom}: $e_{b} = 0.6, e_{c} = 0.8$. 
The left and right panels correspond to $\alpha = 1$ and  
$\alpha = 3/2$, respectively.  
\label{fig:valid}}
\end{center}
\end{figure}

Equation (\ref{eqn:gen}) is valid only for $e^{2}_{\sigma} \ll 1$.  
Jing \& Suto (2002), however, showed that the axes ratios of cluster
scale halos are typically $0.4 < c/a \le b/a <0.8$, which corresponds to 
$0.3 < e_{b} \le e_{c} < 0.8$. To examine how well equation 
(\ref{eqn:gen}) works in this realistic range of $e_{\sigma}$, 
we integrate equation (\ref{eqn:phi1}) numerically to obtain the real 
gravitational potential. Figure \ref{fig:valid} compares the 
perturbative potentials (dashed lines) to the numerical results  
(solid lines) for various values of the halo eccentricities both in 
$\alpha = 1$ and $\alpha =3/2$ cases.  It is clear that the perturbative 
potentials agree with the numerical results excellently. We also calculate 
the ratios of the perturbative potentials to the numerical ones, and 
find the ratios less than $1.1$ even for $e_{b}= 0.6$ and $e_{c} = 0.8$, 
which  indicates that equation (\ref{eqn:gen}) is indeed a good
approximation. 

Incidently, the gravitational potentials for the axis symmetric halos 
with the single eccentricity $e$ can be also obtained from equation 
(\ref{eqn:gen}): 
\beq
\label{eqn:sym}
\Phi(\bu) \approx C\left[ F_{1}(u) \pm \frac{e^{2}}{2} 
\{F_{2}(u) + \cos^{2}\theta F_{3}(u)\}\right],
\eeq
where the positive and the negative signs in front of $e^{2}/2$ correspond 
to the oblate  [$a = b > c$, $e = \sqrt{1 - (c/a)^{2}}$)] and the 
prolate [$a < b = c$, $e = \sqrt{1 - (a/b)^{2}}$] cases, respectively.  
Note that $z$-direction is always chosen as the symmetric axis in the 
above expression. 

\subsection{Iso-Potential Surfaces and Potential Profiles} 

\begin{figure}
\begin{center}
\plotone{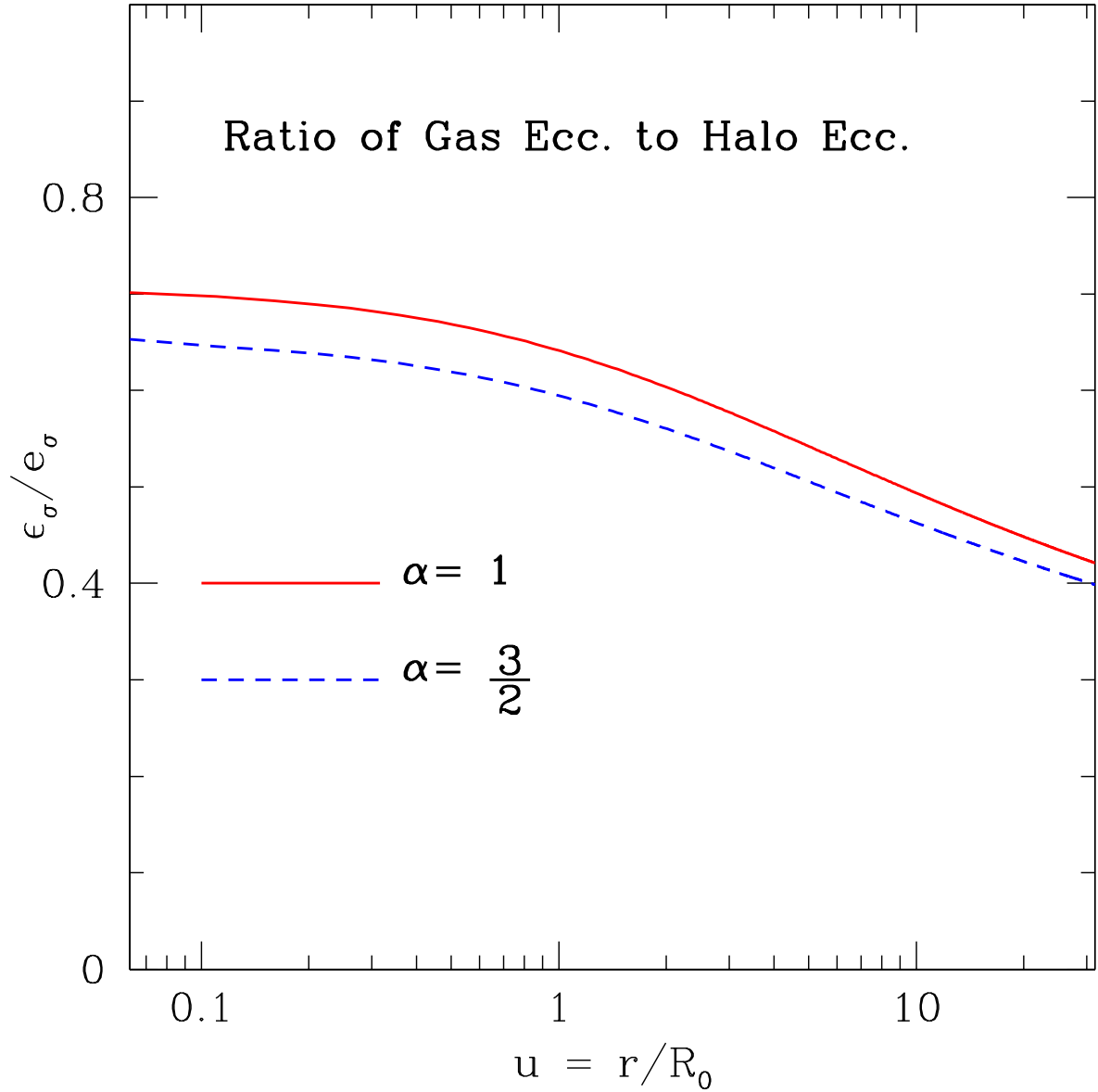}
\caption{The ratio of the eccentricity of the iso-potential surface 
to that of the halo iso-density surface from the perturbative 
result (\ref{eqn:ecc}); $\alpha = 1$ 
(solid line) and $\alpha = 3/2$ (dashed line).
\label{fig:ecc}}
\end{center}
\end{figure}

Strictly speaking, the iso-potential surfaces of ellipsoidal dark halos 
are not necessarily exact ellipsoids \citep{bin-tre87}.  However, 
we have found that the iso-potential surfaces are still best 
well approximated as ellipsoids 
(see also Figs.\ref{fig:ell} and \ref{fig:ell2} below).  
Thus, we  model the iso-potential surfaces as triaxial ellipsoids with 
the rescaled major axis length, $\xi$, and the two eccentricities, 
$\ep_{b}$ and $\ep_{c}$: 
\beq
\label{eqn:isopot}
\xi^{2} \equiv \frac{1}{R^{2}_0}\left(x^2 + \frac{y^2}{1-\ep_b^2} 
+ \frac{z^2}{1-\ep_c^2}\right) = const.   
\eeq
If equation (\ref{eqn:isopot}) is a good and consistent approximation
for the real iso-potential surfaces in the frame of our perturbative 
approach,  we should be able to find the potential profile,
$\tilde{\Phi}$, that satisfies $\Phi(\bu) = \tilde{\Phi}(\xi)$.   
To find a functional form of $\tilde{\Phi}$, we first expand equation  
(\ref{eqn:isopot}) assuming $\ep_{\sigma}^{2} \ll 1$: 
\beq
\label{eqn:isopot2}
\xi= u \left(1 + \frac{\ep_b^2 y^2 + \ep_c^2 z^2}{2r^2}\right) 
= u \left(1 + \frac{\ep_{b}^{2}\sin^{2}\theta\sin^{2}\phi 
+  \ep_{c}^{2} \cos^{2}\theta}{2} \right)  .   
\eeq 
Substituting equation (\ref{eqn:isopot2}) into $\tilde{\Phi}(\xi)$, and 
expanding to firstorder of $\ep^{2}_{\sigma}$, we have    
\begin{eqnarray}
\label{eqn:potf}
\tilde\Phi(\xi) &\approx&  
\tilde\Phi\left[u \left(1 + \frac{\ep_{b}^{2}\sin^{2}\theta\sin^{2}\phi 
+  \ep_{c}^{2} \cos^{2}\theta}{2} \right)\right] \nonumber\\
&=& \tilde\Phi(u) + \frac{\partial \tilde\Phi(u)}{\partial u}
 u  \frac{\ep_{b}^{2}\sin^{2}\theta\sin^{2}\phi 
+  \ep_{c}^{2} \cos^{2}\theta}{2} .
\end{eqnarray}
Comparing equation (\ref{eqn:gen}) with equation (\ref{eqn:potf}) 
indicates that the functional form of $\tilde{\Phi}$ is   
\beq
\tilde\Phi(\xi) = C 
\left[F_{1}(\xi) + \frac{e_{b}^{2}+e_{c}^2}{2}F_{2}(\xi) \right] ,   
\label{eqn:potpfile}
\eeq
and that the eccentricities of iso-potential surfaces
$\ep^{2}_{\sigma}$ are written in terms of their halo eccentricities 
$e^{2}_{\sigma}$ as  
\beq
\label{eqn:ecc}
\frac{\ep^{2}_{\sigma}}{e^{2}_{\sigma}} = 
\frac{F_{3}(u)}{u\partial_{u}F_{1}(u)} = 
\frac{(\alpha -2)F_{3}(u)}{1 - (\alpha -2)F_{1}(u) 
- u^{2-\alpha}(1+u)^{\alpha-2}}.
\eeq
Note that $\ep_{\sigma}$ for the gravitational potential depends on $u$ 
unlike the constant $e_{\sigma}$ for the adopted dark matter halo profile.   
Figure \ref{fig:ecc} plots $\ep_{\sigma}/e_{\sigma}$ as a function 
of $u$ for $\alpha = 1$ and $\alpha = 3/2$ cases. 
The $\alpha = 1$ case has a higher ratio of $\ep_{\sigma}$ to
$e_{\sigma}$ than the $\alpha = 3/2$ case. 
In the whole range of $u$, $\ep_{\sigma}/e_{\sigma}$ is less than unity, 
and decreases mildly as $u$ increases.  

\section{GAS DENSITY AND TEMPERATURE PROFILES} 

In $\S 2$, we showed that the gravitational potential of dark halos 
is the central quantity for the hydrostatic intra-cluster gas  
distribution, and that the gas iso-density and temperature surfaces 
coincide with the halo iso-potential surfaces. Thus, one can restate 
every feature of the iso-potential surfaces found in $\S 3$ in terms 
of the iso-density surfaces of the intra-cluster gas: the 
iso-density and temperature surfaces of the intra-cluster gas in triaxial 
dark halos are approximately triaxial ellipsoids whose eccentricities 
are related to that of the underlying halos via equation
(\ref{eqn:ecc}), decreasing with radial distance. 

Note that we have derived equation (\ref{eqn:ecc}) using three different 
approximations: the first order approximation of the gravitational 
potential with the perturbation expansion (eq.[\ref{eqn:gen}]);
the approximation of the iso-potential surfaces with triaxial 
ellipsoids (eq.[\ref{eqn:isopot}]); the first order approximation 
of the iso-potentential surface eccentricities with the perturbation 
expansion (eq.[\ref{eqn:potf}]). The accuracy of equation (\ref{eqn:ecc}) 
depends on the accumulated errors from the three approximations.  
In $\S 3.2$, we have already shown that equation (\ref{eqn:gen}) is 
an excellent approximation within the $10 \%$ error.
 
\begin{figure}
\begin{center}
\plotone{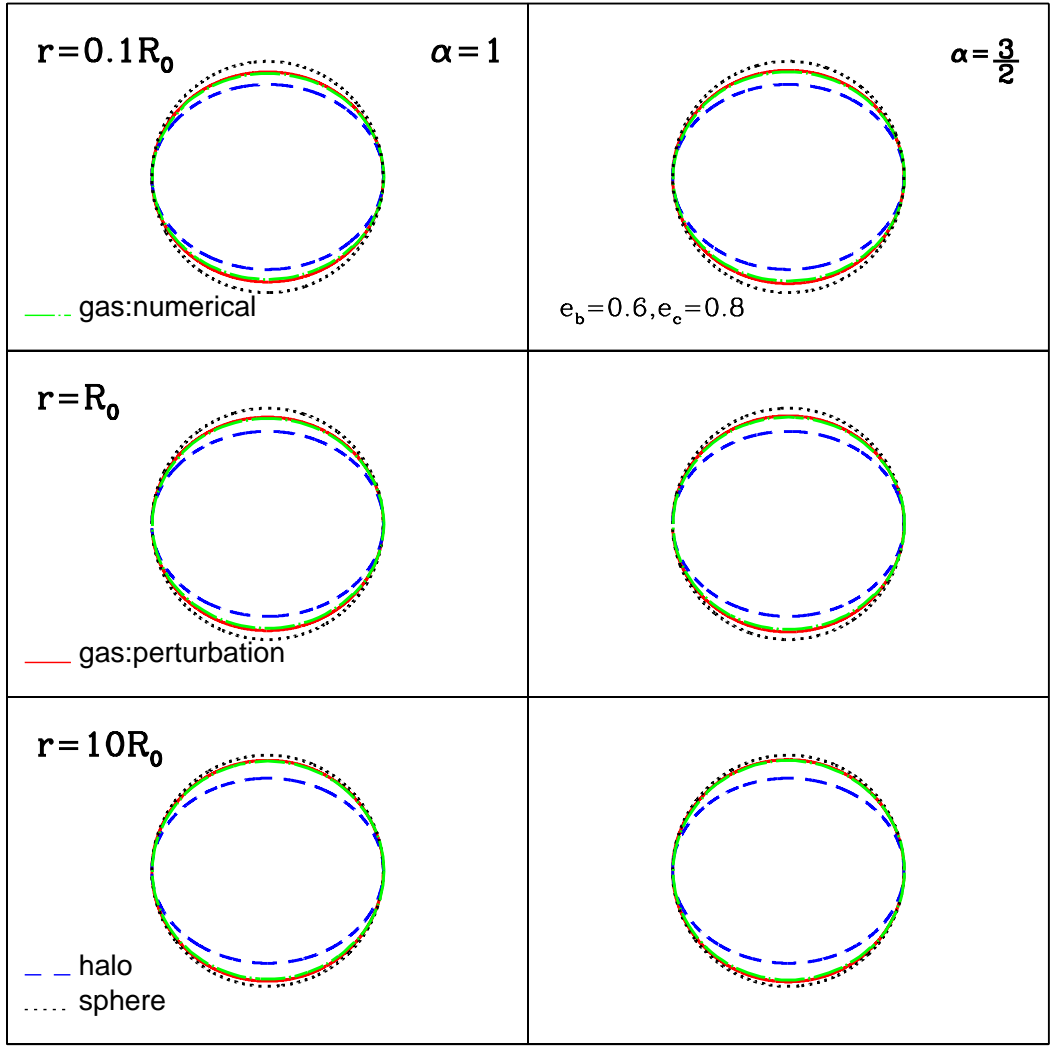} 
\caption{The shapes of the iso-density surfaces of the gas 
embedded in the halo with the eccentricities $e_{b} = 0.6$ and $e_{c} = 0.8$
at three different radii, $r = 0.1R_{0},R_{0}$ and $10R_{0}$ 
on the equatorial plane ($\theta = \pi/2$). The sizes of the curves 
are arbitrary.  
In each panel, the solid and the dot-dashed lines represent the 
the numerical and the perturbative results of the gas density profile, 
respectively. While the dashed lines represents the halo iso-density 
surfaces. The dotted circles are also plotted for comparison. 
\label{fig:ell}}
\end{center}
\end{figure}

Figure \ref{fig:ell} plots the 2D contours of the gas iso-density 
surfaces (solid lines) and the ellipsoidal approximations (dot-dashed lines) 
on the equatorial plane ($\theta = \pi/2$) at three different radii  
$r = 0.1R_{0}, R_{0}$ and $10R_{0}$ for the given halo eccentricities 
$e_{b} = 0.6$ and $e_{c} = 0.8$. The contours of the gas iso-density 
surfaces are found numerically directly from the gravitational potential 
given as equation (\ref{eqn:phi1}).  We also plot the 2D shapes 
of the underlying halos (dashed lines) on the equatorial plane, and 
the circles (dotted lines) as well for reference.
Figure \ref{fig:ell} shows that the approximation of the gas density surfaces 
(or equivalently,the iso-potential surfaces) as ellipsoids work quite
well even for large halo eccentricities $e_{b} = 0.6$ and $e_{c} = 0.8$. 
It also demonstrates explicitly that the intra-cluster gas is more 
spherical in the outer part than in the inner part of the potential, 
and that it is overall more spherical than the underlying dark halo. 
This is intuitively understood because the potential represents the 
overall average of the local density profile, and also because the 
gas pressure is isotropic unlike the possible anisotropic velocity 
ellipsoids for collisionless dark matter. 
This fact has been also detected by hydrodynamic simulations 
\citep[e.g., see Fig. 4 in][]{yos-etal01}. 
 
Figure \ref{fig:ell2} plots the same as Figure \ref{fig:ell} but for 
the case of halo eccentricities, $e_{b}=0.8$ and $e_{c}=0.8$. 
It shows clearly that the ellipsoidal approximations (dot-dashed lines) 
still work fairly well but that there are some noticeable difference  
between the perturbative (solid lines) and the numerical (dot-dashed lines)  
results.  To quantify the difference of the ellipsoidal approximations 
of equation (\ref{eqn:ecc}) to the gas iso-density surfaces, 
we measure the eccentricities of the gas iso-density surfaces directly 
from the contours of the numerically calculated gravitational potential, 
and determine the ratios between these numerical eccentricities 
$\epsilon^{num}_{\sigma}$ and analytical approximations   
$\epsilon_{\sigma}^{per}$. Figures \ref{fig:dev1} and \ref{fig:dev2}
show the results as functions of the halo eccentricities.  
Note that $\epsilon_{b}^{num}$ depends on both $e_{b}$, and $e_{c}$, 
so that $\epsilon_{b}^{num} \ne \epsilon_{b}^{per}$ even when $e_{b} =
0$ if $e_{c} \ne 0$. In contrast, for the analytical approximations of 
equation (\ref{eqn:ecc}), $\epsilon_{b}$ depends only on $e_{b}$. 
Figures \ref{fig:dev1} and \ref{fig:dev2} also suggest that 
$\epsilon_{b}^{num}/\epsilon_{b}^{per}$ is not so sensitive to 
the values of $u$ and $\alpha$. On the whole, the values of 
$\epsilon_{b}^{num}/\epsilon_{b}^{per}$ are less than $1.2$, indicating 
that the error involved in approximating $\epsilon_{b}^{num}$ to 
$\epsilon_{b}^{per}$ is less $20 \%$ at most. Furthermore, 
to the approximation error, we provide the following fitting 
formula to $\epsilon_{b}^{num}/\epsilon_{b}^{per}$: 
\ben
\frac{\epsilon_{b}^{num}}{\epsilon_{b}^{per}} 
&=& 1 + [0.1 + 0.05\log(1+u)]e_{c}^{3} + 
[0.2 + 0.03\log(1+u)]e_{b}^{3}, \label{eqn:fit1}\\ 
\frac{\epsilon_{c}^{num}}{\epsilon_{c}^{per}} 
&=& 1 + [0.1+0.09\log(1+u)]e_{b}^{3} + [0.2+0.03\log(1+u)]e_{c}^{3}.
\label{eqn:fit2}
\een
Equations (\ref{eqn:fit1}) and (\ref{eqn:fit2}) represent the accuracy 
of equation (\ref{eqn:ecc}), quantifying the calculation errors accumulated 
from all the above three approximations made in the derivation of 
equation (\ref{eqn:ecc}). In summary, equation (\ref{eqn:ecc}) is accurate 
within $10\%$ errors for $e_{\sigma} < 0.6$, while within $20\%$ errors  
for $e_{\sigma} < 0.8$. 

By equations (\ref{eqn:iso}),(\ref{eqn:pol}) and (\ref{eqn:gen}), 
we obtain the density and temperature profiles for both the isothermal and 
the polytropic gases in the triaxial dark halos, and plot the final
results in Figure \ref{fig:profile} for $\alpha = 1$ and $\alpha = 3/2$ 
cases.  For this figure, we choose $e_{b} = 0.6, e_{c}=0.8$, $\gamma = 1.15$, 
and $(1-\gamma)/(K_{0}\gamma) = 1$.  For comparison, we also plot 
the axis-symmetric (both oblate and prolate with $e = 0.8$) 
and the spherical cases. 

Both in isothermal and polytropic cases, the significant deviation 
of the resulting profiles from the sphericity are manifest especially 
in the range of $r \ge R_{0}$, increasing with $r$.  
The $\alpha = 3/2$ case shows non-negligible deviations even for 
$r \le 10^{-1}R_{0}$. It is interesting to note that the oblate 
and prolate profile curves are symmetric about the spherical ones  
This is due to the fact that the gravitational potentials for the oblate 
and the prolate halos differ only by the sign before the non-spherical 
perturbative term when the $z$-direction is chosen as the symmetric axis 
(see $\S 3.1$).  

\section{CONCLUSIONS AND DISCUSSION}

We have adopted the triaxial halo density profile suggested by 
Jing \& Suto (2002), and calculated the halo gravitational potential 
by using perturbation, assuming the small eccentricities of the 
underlying halo. The approximations have been shown to be valid  
even for the halos with fairly large eccentricities. 
With the resulting halo potentials, we found the solutions to the hydrostatic 
equilibrium gas equations both for the isothermal and polytropic gases. 
The corresponding gas density and temperature profiles have been 
shown to deviate from the conventional spherical models significantly 
at large radial distances. 

We have also derived a useful analytical formula for the eccentricity ratio 
between the gas and the halo as a function of radial distance. 
It has been shown that the intra-cluster gas is rounder than 
the underlying halo, and that the gas eccentricity decreases with radial
distance. It provides a quantitative explanation about why the gas inside
dark matter halos are observed to be less elongated than the halo 
themselves in hydrodynamic simulations. 
We expect wide applications of this formula: First, it can be 
applied to clusters to determine the shapes of the
unseen dark halos from the observed shapes of the intra-cluster 
gas from the X-ray and Sunyaev-Zel'dovich effects. Second, 
it can be used to reduce the errors involved in the measurement of 
the Hubble constant caused by the cluster asphericity. Third,  
it can be also useful in determining cosmic shear from the 
weak gravitational lensing surveys, and so on. 

It is worth emphasizing that our analytical results are not empirical 
fitting formula, but derived from the first principles using the 
perturbation theory.  The only assumptions made in our 
derivation is that the intra-cluster gas is in hydrostatic equilibrium. 
Thus, it will be also possible to test the validity of hydrostatic 
equilibrium of the intra-cluster gas model using our analytical results.
Furthermore, our results are quite general in the sense that 
they are derived for any arbitrary value of the gas constants,  
$\gamma$ and $K_{0}$, and thus can used to constrain 
the values of $\gamma$ and $K_{0}$ through the direct comparison 
with the observational data. 
 
\acknowledgments

We thank A. Taruya, M. Oguri, and T. Kuwabara for their useful
comments. We also thank the anonymous referee for careful reading of 
the original manuscript and many helpful suggestions.   
J. L. gratefully acknowledges the support from the JSPS fellowship. 
This research was supported in part by the Grant-in-Aid for Scientific 
Research of JSPS (12640231, 14102004, 14-02038).

\appendix

\section{Analytic Expressions for $\alpha=1$ and $\alpha=3/2$ cases}

For the inner spectral indices of $\alpha=1$ and $\alpha=3/2$, which are
indeed suggested from numerical simulations, 
$F_i$ and $\epsilon_{\sigma}/e_{\sigma}$ defined in \S 3 can be 
written explicitly in terms of elementary functions. 
We provide their expressions here which may be useful in quantitative
confrontation of our results with observations.

For $\alpha = 1$, \\ 
\ben
F_{1}(u) &=& -\frac{1}{u}\ln(1 + u), \\ 
F_{2}(u) &=& -\frac{1}{3} + \frac{2u^2 - 3u + 6}{6u^2} + 
\left(\frac{1}{u} - \frac{1}{u^{3}}\right)\ln(1 + u),\\ 
F_{3}(u) &=& \frac{u^{2} -3u - 6}{2u^{2}(1+u)} + \frac{3}{u^{3}}\ln(1+u),  
\een
\beq
\frac{\ep^{2}_{\sigma}}{e^{2}_{\sigma}} = 
\frac{6(1+u)\ln(1+u) + u^{3}-3u^{2}-6u}{2u^{2}[(1+u)\ln(1+u)-u]}.
\eeq

For $\alpha = 3/2$, \\ 
\ben
F_{1}(u) &=& -2  + 2\sqrt{\frac{1+u}{u}}
- \frac{2}{u}\ln(\sqrt{u} + \sqrt{1+u}), \\
F_{2}(u) &=& \frac{4}{3} + 
\left(\frac{5}{4u^{3}}+\frac{5}{12u^{2}}-\frac{13}{6u}
- \frac{4}{3}\right)\sqrt{\frac{u}{1+u}} + 
\left(\frac{2}{u} - \frac{5}{4u^{3}}\right)\ln(\sqrt{u} + \sqrt{1+u}), \\ 
F_{3}(u) &=&  -\left(\frac{15}{4u^{3}}+\frac{5}{4u^{2}}-\frac{1}{2u}\right)
\sqrt{\frac{u}{1+u}} + \frac{15}{4u^{3}}\ln (\sqrt{u} +\sqrt{u+1}),  
\een

\beq
\frac{\ep^{2}_{\sigma}}{e^{2}_{\sigma}} = 
\frac{\left(15-5u-2u^{2}\right)\sqrt{u/(1+u)} - 15\ln(\sqrt{u} + \sqrt{1+u})}
{8u^{3}\sqrt{u(1+u)} - 8u^{2}\ln(\sqrt{u} + \sqrt{1+u})}.
\eeq

\clearpage

\clearpage

\begin{figure}
\begin{center}
\plotone{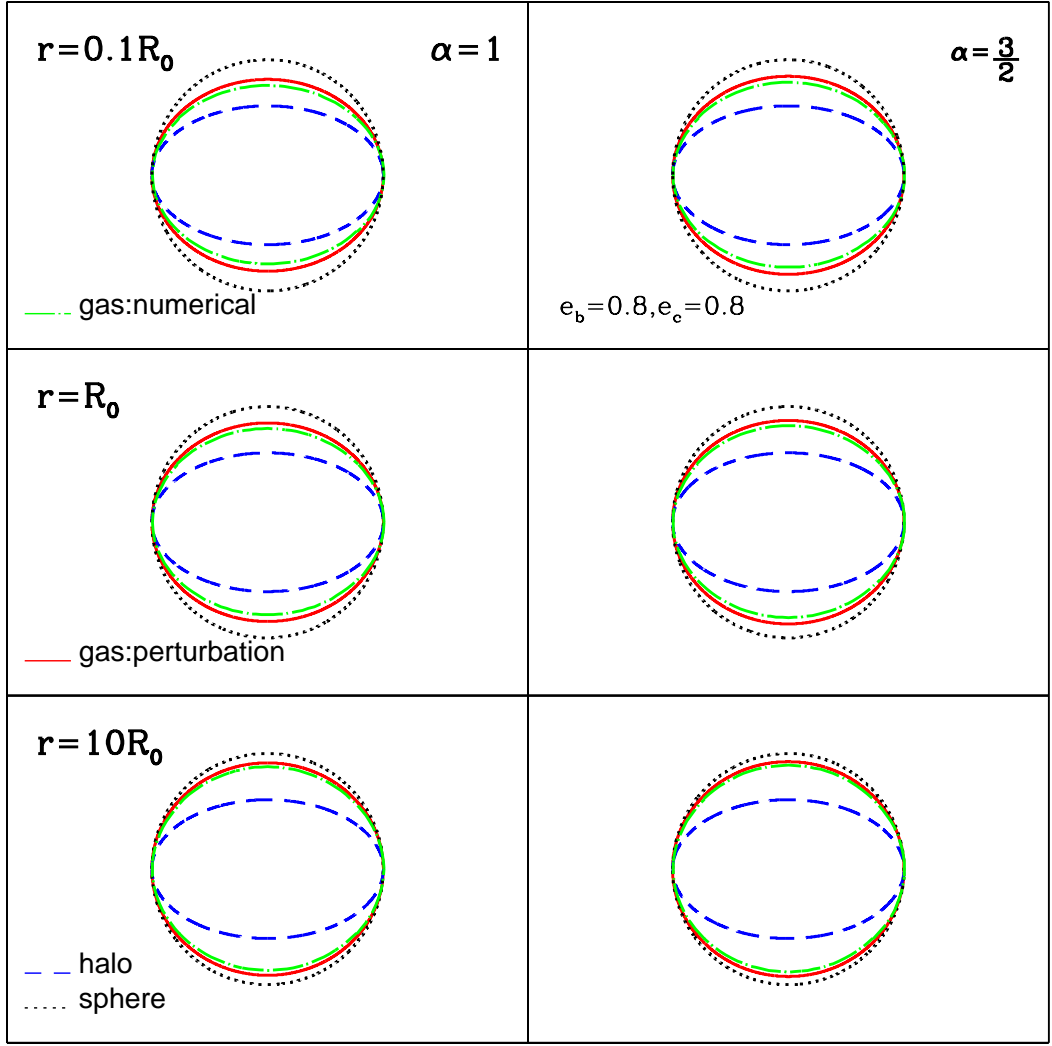} 
\caption{Same as Figure \ref{fig:ell}, but for larger halo 
eccentricities,$e_{b} = 0.8$ and $e_{c} = 0.8$.   
\label{fig:ell2}}
\end{center}
\end{figure}

\begin{figure}
\begin{center}
\plotone{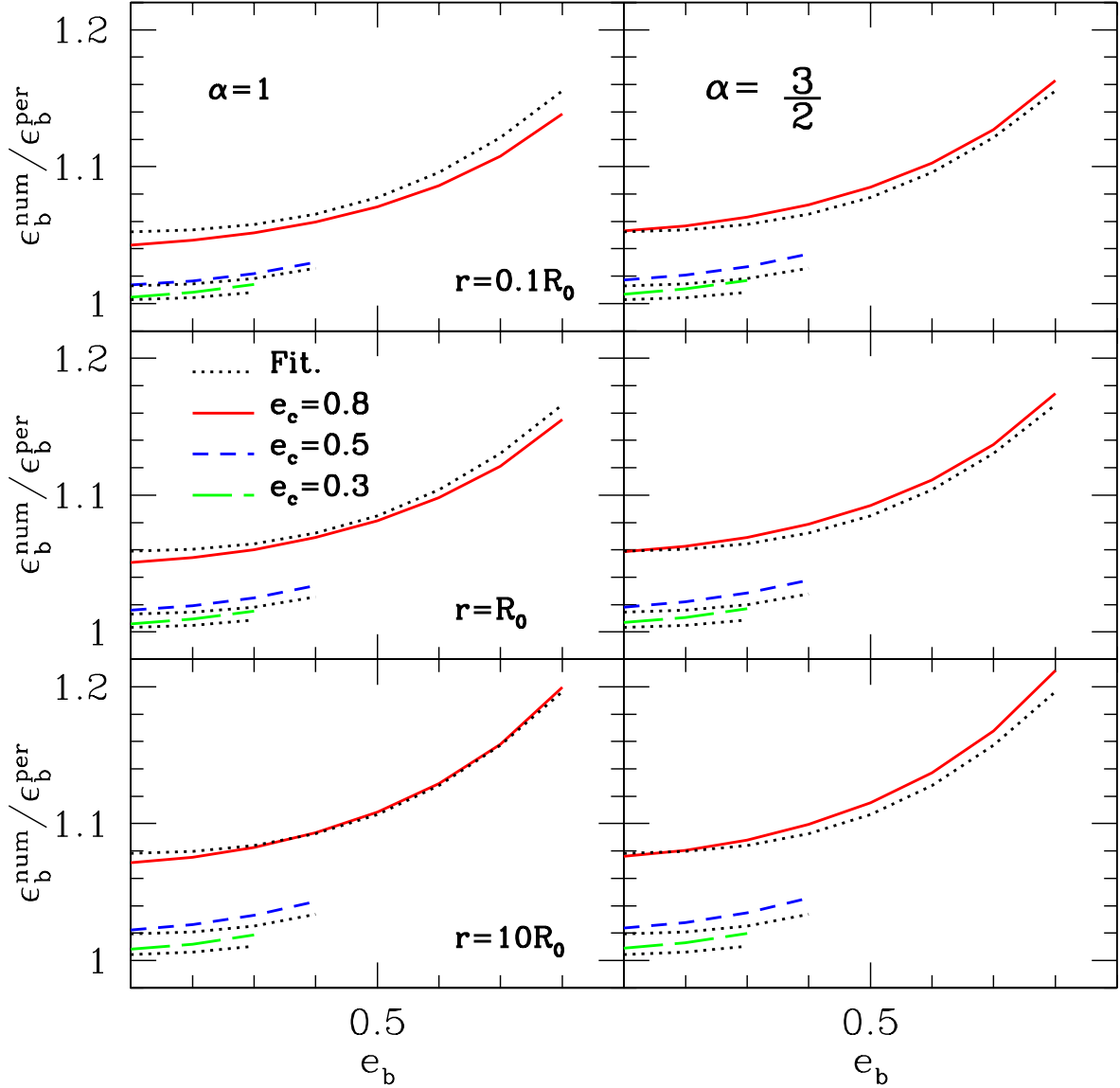} 
\caption{The ratio between the gas eccentricity fitted to the numerical 
results and the perturbative ones as a function of the halo eccentricity 
$e_{b}$. In each panel, the solid, the dashed and the long-dashed lines 
represent the cases of $e_{c} = 0.8,0.5$ and $0.3$, respectively. 
The dotted line represents the fitting formula given in equations 
(\ref{eqn:fit1}) and (\ref{eqn:fit2}).  
\label{fig:dev1}}
\end{center}
\end{figure}

\begin{figure}
\begin{center}
\plotone{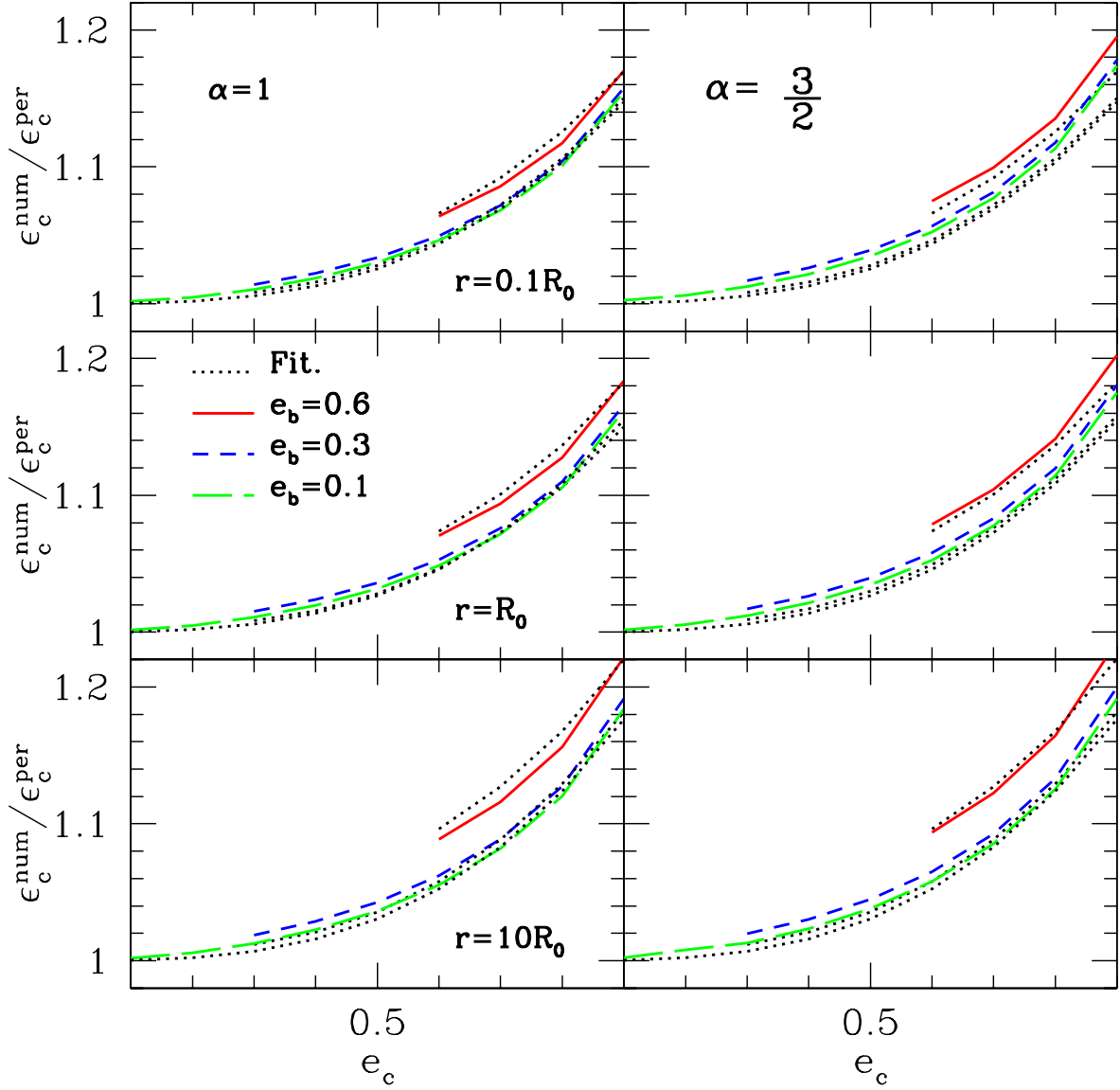} 
\caption{Same as Figure \ref{fig:dev1}, but as  
a function of $e_{c}$ for the three different values of 
$e_{b} = 0.1, 0.3$ and $0.6$.  
\label{fig:dev2}}
\end{center}
\end{figure}

\begin{figure}
\begin{center}
\plotone{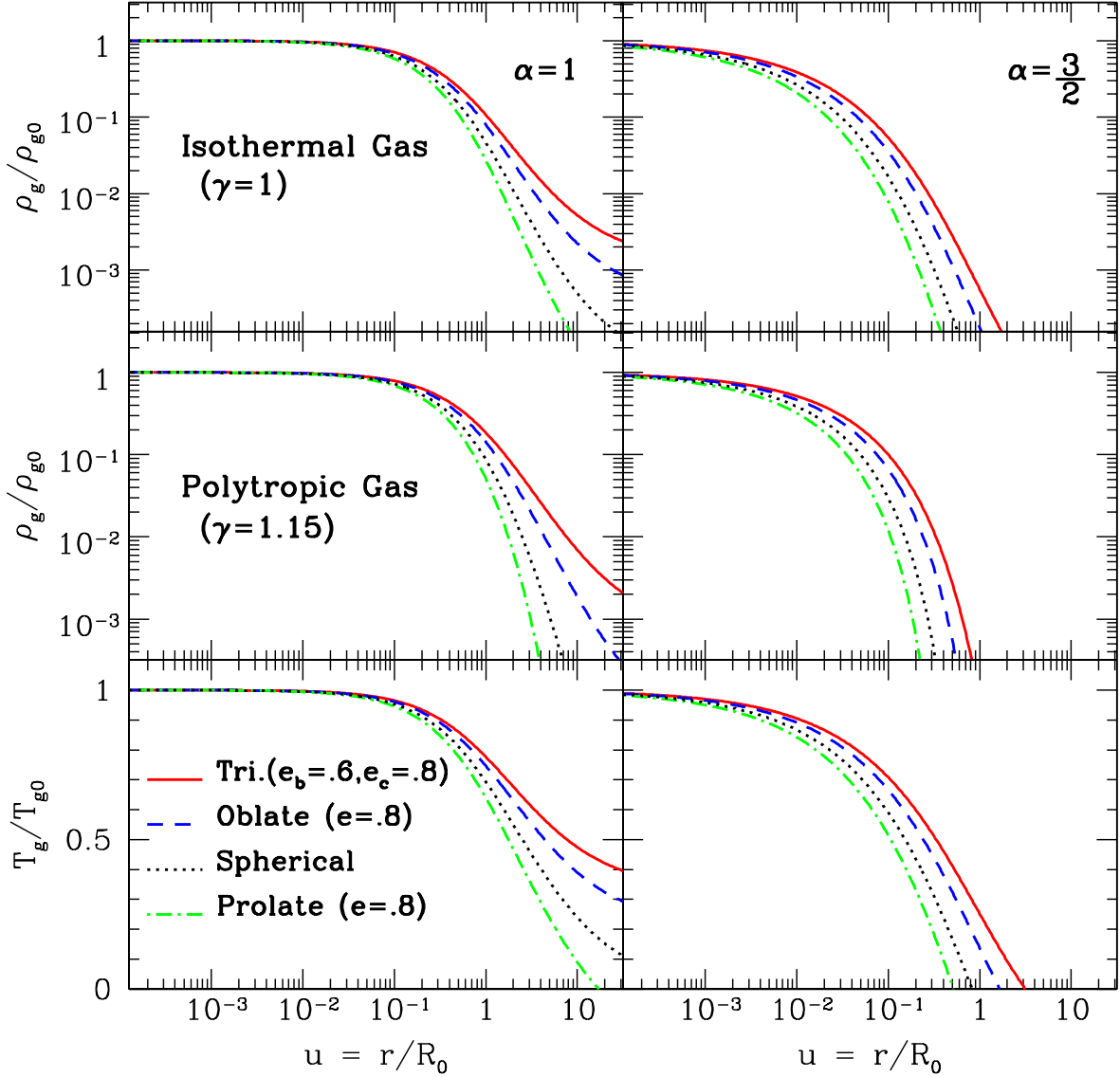} 
\caption{The hydrostatic equilibrium solutions for the intra-cluster gas
distribution. 
{\it Top}: the density profiles of the isothermal gas. 
{\it Middle}: the density profiles of the polytropic gas with the 
polytropic index of $\gamma = 1.15$.   
{\it Bottom}: the temperature profiles of the polytropic gas. 
In each panel, the solid, the dashed, the dotted, and the dot-dashed 
lines represent the gas profiles for the 
triaxial ($e_{b}=0.6$, $e_{c}=0.8$), the oblate ($e=0.8$), the spherical, 
and the prolate ($e=0.8$) halo cases, respectively. 
\label{fig:profile}}
\end{center}
\end{figure}

\end{document}